\documentclass[sigconf]{acmart}
\renewcommand\footnotetextcopyrightpermission[1]{} 

\usepackage{lingmacros}
\usepackage{tree-dvips}
\usepackage{amsmath}
\usepackage{amsfonts}
\usepackage{amsthm}
\usepackage{hyperref}
\usepackage{subcaption}
\newtheorem{theorem}{Theorem}
\newcommand{\myvec}[1]{\ensuremath{\mathbf{{#1}}}}   
\newcommand{\mymat}[3]{\ensuremath{\mathbf{#1}^{#2}_{#3}}}        
\newcommand{\myhref}[3][blue]{\href{#2}{\color{#1}{#3}}}        

\AtBeginDocument{%
  \providecommand\BibTeX{{%
    \normalfont B\kern-0.5em{\scshape i\kern-0.25em b}\kern-0.8em\TeX}}}

\setcopyright{none} 
\begin{document}

\title{Diversifying Relevant Phrases}







\author{Shreya Malani, Dinesh Gaurav, Anoop Vallabhajosyula, Rahul Agrawal}
\affiliation{\institution{Microsoft India Development Center}}
\email{{shreyam, digaurav, anvallab, rahulagr}@microsoft.com}






\begin{abstract}
Diverse keyword suggestions for a given landing page or matching queries to diverse documents is an active research area in online advertising. Modern search engines provide advertisers with products like Dynamic Search Ads and Smart Campaigns where they extract meaningful keywords/phrases from the advertiser’s product inventory. These keywords/phrases are representative of a diverse spectrum of advertiser’s interests. In this paper, we address the problem of obtaining relevant yet diverse keywords/phrases for any given document. We formulate this as an optimization problem, maximizing the parameterized trade-off between diversity and relevance constrained over number of possible keywords/phrases. We show that this is a combinatorial NP-hard optimization problem. We propose two approaches based on convex relaxations varying in complexity and performance. In the first approach, we show that the optimization problem reduces to an eigen value problem. In the second approach, we show that the optimization problem reduces to minimizing a quadratic form over an $\mathit{l}_1$-ball. Subsequently, we show that this is equivalent to a semi-definite optimization problem. To prove the efficacy of our proposed formulation, we evaluate it on various real-world datasets and compare it to the state-of-the-art heuristic approaches.



\end{abstract}


\maketitle
\pagestyle{plain} 

\section{Introduction}
Online advertising is a $100$+ billion dollars market that forms the backbone of Internet’s economics. Sponsored search is the most prominent kind of online advertising, where ads are shown in response to search queries. In its most common form, a well-directed advertising setup requires multiple inputs from the advertisers - keywords which describe the product being advertised, budget that the advertiser is willing to spend on each advertisement click and the advertisement itself (with all the attractive terms that describe the product). Most relevant keywords/phrases are a representation of their possibly diverse product portfolio. The search engine then maps these relevant keywords/phrases to matching queries issued by users of the search engine and displays ads along with the search results. Advertisers pay to the search engines when the users click on the displayed ads. Since the ads are shown along with the search results, it is extremely critical to show relevant ads, to ensure good user experience and sustain continued trust in the quality of the underlying search engine. Due to the rapid growth and proliferation of internet-based economics, most advertisers need to have a strong online presence covering two important aspects - a complete online inventory describing their diverse portfolio which is typically their website and a very well set-up campaign in the publisher/search engine's platform.

While the larger advertisers can afford to invest in research and engineering efforts to manage their online campaigns, this is far more difficult for small and medium sized advertisers (SMB). This has prompted the search engines to build advertiser-friendly tools to automate the campaign generation and maintenance process. These automation tools have been key drivers to increase investments by SMBs. 
 With the launch of \myhref{https://support.google.com/google-ads/answer/2471185?hl=en
}{Dynamic Search Ads (DSA)}, \myhref{https://support.google.com/google-ads/answer/2375499?hl=en}{Ad Extensions}, \myhref{https://support.google.com/google-ads/answer/7652860?hl=en&ref_topic=9024773}{Smart Campaigns (SC)}, and \myhref{https://support.google.com/google-ads/answer/6325042?hl=en}{Automated Bidding}, search engines are trying to address the problem of generating relevant keywords/phrases for the advertiser without any effort from the later. For instance, typically in a product like DSA, advertisers share their website links and their daily budget on the advertising campaign. Subsequently, the search engine does the following 
\begin{itemize}
\item The search engine crawls the web inventory of the advertiser and generates a collection of relevant keywords/phrases. Keyword suggestions from web page (inventory) is a well-studied problem. For instance, Rapid Keyword Extraction (RAKE), proposed in \cite{rose2010automatic}, is a well-known algorithm for keyword suggestions from large texts on landing pages. For further references, also see \cite{siddiqi2015keyword}. 
\item The search engine then maps these keywords/phrases to queries from their own search history. For this, both the keywords/phrases and queries are mapped into vector representations. Consequently, for each query, the search engine tries to find the most relevant keywords/phrases, typically using methods falling under the umbrella of Neighborhood Graph Search (NGS) \cite{wang2012query}.
\end{itemize}

While the above approaches focus solely on retrieving relevant keywords/phrases there has been a parallel line of research which also tries to diversify the suggested keywords/phrases for a given user query. For example, the advertiser who is a re-seller for apple (electronic industry) products would be interested in the query "apple" and would like to match it to her diverse set of apple products. Thus, the advertiser would prefer to emphasize diversity over relevance in this particular scenario. This necessitates a parameterized way of achieving a trade-off between diversity and relevance. Often these are formulated as difficult combinatorial optimization problems which are then solved using simple and intuitive heuristic approaches. We do a brief survey on this in a later section.

In this paper, following similar lines as earlier works, we formulate an optimization problem for maximizing the trade-off between relevance and diversity constrained over the count of possible candidate suggestions. Our contributions are as follows,
\begin{itemize}
\item We propose two approaches based on convex relaxations to solve the formulated combinatorial optimization problem. In the first approach, we show that the optimization problem reduces to finding eigen values of a symmetric matrix. In the second approach, we show that the same can be relaxed to a standard semi definite optimization problem.
\item We extensively evaluate our proposed approaches on real world datasets which are publicly available. We also compare them with popularly known heuristic algorithms in literature for this problem. In addition, we do the same on actual data from a well-known search engine.
\end{itemize}

The remaining part of the paper is divided into related work, followed by problem formulation and proposed solutions. The later sections give the details on the experiments and the results. 

\section{Related Work}

One of the earlier known works on diversification of query results is Maximal Marginal Relevance (MMR) \cite{Carbonell:1998:UMD:290941.291025}. The authors propose MMR ranking to order the retrieved documents for search results. The ranking is a linear combination of the query and document similarity and the document-to-document similarities. In successive iterations, MMR updates the candidate set by selecting the current best candidate that maximizes the objective with respect to the current candidate set. The optimization objective represents the trade-off between relevance and diversity and is controlled by a parameter. The algorithm stops when it finds the candidate set of required cardinality. 
 
In \cite{Jain03providingdiversity} paper, the authors propose the algorithm referred as Motley, which is an iterative approach of picking next most similar point whose diversity with points in current candidate set is above a threshold. Whereas Bswap, described in \cite{Yu09ittakes}, starts with the top $k$ relevant points and keeps swapping elements from this set while ensuring increase in diversity and decrease in similarity is under a threshold.

A thorough experimentation and comparison of the available diversification algorithms is explained in \cite{DBLP:conf:icde:VieiraRBHSTT11}. The authors propose an iterative approach Greedy Marginal Contribution (GMC) which enhances the results set so far by adding an item which increases diversity of the results set. In the second algorithm Greedy Randomized with Neighborhood Expansion (GNE) authors allow swapping of the elements in the result set apart from adding only to achieve more optimal solution. 
We extensively compare our proposed approaches to above mentioned algorithms.

\section{Problem Statement}
Our notations closely follow the lines of \cite{DBLP:conf:icde:VieiraRBHSTT11}. Let $U = \{s_1,\dots,s_n\}$ be the universal set of relevant candidate keyword/phrases with cardinality $n$ and $q$ be the user query. 
We denote $\delta_{sim}({q,s_i})$ as any similarity function between the vector representations of $q$ and $s_i$. Intuitively, this function will be a non-decreasing function of the similarity between them. In practice, this can even be the click-through-rate that query $q$ received when matched to the keyword $s_i$ in the context of sponsored search. Similarly, we denote $\delta_{div}({s_i,s_j})$ as an abstract metric that represents the distance between two keywords $s_i$ and $s_j$. For instance, in context of vector representations, this can be the geometric distance between the keywords (euclidean, cosine \dots). 



We define a subset $S = \{s_{i_1},\dots,s_{i_k}\}$ where $\{i_1,\dots,i_k\}$ are $k$ indices from the set $U$. We formulate the optimization problem of finding the optimal $S$ that maximizes the trade-off between the relevance and diversity of the chosen candidates by defining the objective function as
\begin{align}
F(q,S) = (1 - \lambda) * \delta_{sim}(q,S) + \lambda * \delta_{div}(S)
\label{fsum}
\end{align}


Let $x_i$ be the Boolean indicator variable representing the membership of candidate $i$ in $S$. This allows us to re-write the $\delta_{sim}$ and $\delta_{div}$ as
\begin{align}
\delta_{sim}(q,S) &= \sum_{i \in U}x_i*\delta_{sim}(q,s_i) \nonumber \\   
\delta_{div}(S) &= \sum_{i,j \in U}x_i*x_j*\delta_{div}(s_i,s_j)    
\end{align}
Whereas earlier works directly attend to $F(q,S)$ using heuristic approaches in the above form (except for scaling), we slightly reformulate it. We note that $F(q,S)$ is  a non-homogeneous quadratic function owing to the linearity of the relevance function which has linear terms in $x_i$. We propose to replace the linear similarity function by squaring it. While this is motivated by the fact that homogeneous quadratic functions are more amenable to analysis, we also note that the similarity function still retains the monotonicity. In fact, for $\lambda =0$, where diversity is not important, solutions for both the squared and the linear functions are same.  Thus, the objective we try to optimize is given as 
\begin{align}
F^{'}(q,S) &= (1 - \lambda) * (\delta_{sim}(q,S))^2  + \lambda * \delta_{div}(S)  \nonumber \\
&\geq \sum_{i \in U}x_i*x_j*(\delta_{sim}(q,s_i)\delta_{sim}(q,s_j))^{(1-\lambda)} *(\delta_{div}(s_i,s_j))^\lambda \nonumber \\
&= \myvec{x}^T\myvec{A}\myvec{x}
\label{ObjFunc}
\end{align}
where $\myvec{A}$ is the $n\times n$ symmetric matrix such that
\begin{align}
\myvec{A}_{ij} \,=\, (\delta_{sim}^2(q,s_i))^{(1-\lambda)} *(\delta_{div}(s_i,s_j))^\lambda
\end{align}
and $\myvec{x} = [x_1,\dots,x_k]$. The second inequality in \eqref{ObjFunc} follows from application of AM-GM inequality to individual terms. This allows us to formulate the optimization problem of maximizing the trade-off between relevance and diversity as follows

\begin{align}
\max_{\myvec{x} \in \mathbb{R}^N} &~~\myvec{x}^T\mymat{A}{}{}\myvec{x} ~~ \nonumber \\ 
&s.t.~~\lvert\lvert\myvec{x}\rvert\rvert_0 \leq K~~,~~ \myvec{x}_i \in \{0,1\}
\label{l0Opt}
\end{align}

\begin{theorem}
	Optimization problem in \eqref{l0Opt} is NP-Hard.
\end{theorem}

\textit{Proof}: We provide a brief sketch of the proof. We first show that an instance of the above problem is same as the maximum edge weight $K$-clique problem (MEWKCP) \cite{alidaee2007solving}. Further, we note that the maximum edge weight clique problem (MEWC) reduces polynomially to MEWKCP. Since MEWC is an NP-Hard problem, and since an instance of problem under consideration in equivalent to it, problem in $\eqref{l0Opt}$ is NP-Hard.

Consider any completely connected graph $G=(V,E)$. Now, add a node $c$ to $G$ such that $d_{ic}$ is $1$ for all nodes. Thus, form the Graph $G'=(V',E')$ such that $V'(G') = V(G) + c$ and $E'(G') = E(G) + \{(g,c)\lvert g\in V(G)\}$. Substituting $d_{ic}=1$ for all $i$, it is not hard to see that the problem in \eqref{l0Opt} is equivalent to MEWKCP in graph $G$. Since MEWKCP is NP-Hard, this particular instance of the problem is NP-Hard, and thus the problem in general.

In the following subsections, we propose two convex relaxations for the above optimization problem, which majorly has two changes
\begin{itemize}
	\item We replace the $\mathit{l}_0$ constraint with the corresponding $\mathit{l}_2$ and $\mathit{l}_1$ constraints. 
	\item We replace the Boolean constraint with the continuous interval constraint $\myvec{x}_i\in[0,1]$.	
\end{itemize}

We note that there is wide consensus in optimization community that these are often the tightest relaxations and has been theoretically proved in several cases \cite{luo2010semidefinite}.
\subsection{$\mathit{l}_2$-relaxation}
We relax the above problem as the following convex optimization problem
\begin{align}
\max_{\myvec{x} \in \mathbb{R}^N} &~~\myvec{x}^T\mymat{A}{}{}\myvec{x} ~~ \nonumber \\  
&s.t.~~\lvert\lvert\myvec{x}\rvert\rvert_2 \leq K~~,~~0\leq \myvec{x}_i\leq 1
\label{l2_orig}
\end{align}
Note that there are two relaxations here

\begin{theorem}
$\mathit{l}_2$ relaxation has the following solution 
 \begin{align}
 \myvec{x}_i^* = \min \left\{ \myvec{v}_i,\frac{1}{\sqrt{K}}\right\}
 \end{align}
 	where $\myvec{v}$ is the unit-norm eigenvector corresponding to the largest eigenvalue of $\mymat{A}{}{}$.
\end{theorem}
\textit{Proof}: Define $\myvec{y}_i = \frac{\myvec{x}_i}{\sqrt{K}}$. We can re-write the optimization in \eqref{l2_orig} as 
\begin{align}
\max_{\myvec{y} \in \mathbb{R}^N} &~~\myvec{y}^T\mymat{A}{}{}\myvec{y} ~~ \nonumber \\ 
&s.t.~~\lvert\lvert\myvec{y}\rvert\rvert_2 \leq 1~~,~~0\leq \myvec{y}_i\leq \frac{1}{\sqrt{K}}
\label{l2_y}
\end{align}
discarding the constant $K$ in objective as it doesn't change the solution. Consider the following relaxation of this problem which is an upper bound
\begin{align}
\max_{\myvec{y} \in \mathbb{R}^N} &~~\myvec{y}^T\mymat{A}{}{}\myvec{y} ~~ \nonumber \\  
&s.t.~~\lvert\lvert\myvec{y}\rvert\rvert_2 \leq 1
\label{l2_y_relaxed}
\end{align}
From the results on classical matrix analysis, it follows that the solution to above is $\myvec{v}$, the eigenvector corresponding to the largest eigenvalue of $\myvec{A}$. Given that $\myvec{A}$ is also a non-negative irreducible matrix, it also follows that $\myvec{v}_i \geq 0$. Let $\myvec{y}_o$ be the optimal solution to \eqref{l2_y}, it follows that
\begin{align}
\myvec{y}_o^T\myvec{A}\myvec{y}_o \,\leq \, \myvec{v}^T\myvec{A}\myvec{v}
\end{align}
if $\myvec{v}_i \leq \frac{1}{\sqrt{K}},~\forall i$, then required result follows immediately. Let us say it is violated for only one index $\myvec{v}_1$ without loss of generality. Given that $\myvec{A}\geq 0$ and $\myvec{v}_i \geq 0,~\forall i$, one can continuously keep reducing $\myvec{v}_1$ until $\myvec{v}_1 \,=\,\sqrt{K}$ and this should be the optimum for the required optimization problem. Argument for multiple indices violating the constraints follows immediately by induction. 
\subsection{$\mathit{l}_1$-relaxation}
We consider the $\mathit{l}_1$ relaxation
\begin{align}
\max_{\myvec{x} \in \mathbb{R}^N} &~~\myvec{x}^T\mymat{A}{}{}\myvec{x} ~~ \nonumber \\ 
&s.t.~~\lvert\lvert\myvec{x}\rvert\rvert_1 \leq K~~,~~0\leq \myvec{x}_i\leq 1
\label{l1_orig_relaxed}
\end{align}
\begin{theorem}
 $\mathit{l}_1$-relaxation has the following solution 
 \begin{align}
 \myvec{x}_i^* = \min \left\{ \sqrt{\myvec{X}^*_{ii}},\frac{1}{K}\right\}
 \end{align}
where $\myvec{X}^*$ is the positive semi-definite matrix obtained as the solution to the semi-definite optimization problem
\begin{align}
&\max_{\myvec{X} \in \mathbb{R}^{N \times N},\myvec{v}} ~~<\myvec{X},\myvec{A}> ~~ \nonumber \\
&s.t.~~\text{Diag}(\myvec{v})-\myvec{X} \succeq \myvec{0}~~,~~\myvec{e}^T\myvec{v} \leq 1,~~\myvec{X} \succeq 0
\label{X_psd}
\end{align}
\end{theorem}
\textit{Proof}: We skip a complete proof due to the brevity of space and outline a sketch of the same. 
We define the variable $y_i = \frac{x_i}{K}$ which allows us to rewrite the optimization problem as 
\begin{align}
\max_{\myvec{y} \in \mathbb{R}^N} &~~\myvec{y}^T\mymat{A}{}{}\myvec{y} ~~ \nonumber \\ 
&s.t.~~\lvert\lvert\myvec{y}\rvert\rvert_1 \leq 1~~,~~0\leq \myvec{y}_i\leq \frac{1}{K}
\label{l1_y}
\end{align}
It is straightforward to show that solution to above optimization problem can be obtained as $\myvec{y}_i=\min\{\myvec{z}^*_i,\frac{1}{K}\}$. Here $\myvec{z}^*$ is the solution to the optimization problem
\begin{align}
\max_{\myvec{z} \in \mathbb{R}^N} &~~\myvec{z}^T\mymat{A}{}{}\myvec{z} ~~ \nonumber \\ 
&s.t.~~\lvert\lvert\myvec{z}\rvert\rvert_1 \leq 1
\label{l1_y_remove_bounds}
\end{align}
This follows from the fact both the objective and constraints are monotonic in $\myvec{y}_i$. We refer the interested readers to \cite{pinar2006semidefinite} for more details on transforming \eqref{l1_y_remove_bounds} to \eqref{X_psd}. We implement the proposed relaxations using standard convex optimization packages.

\begin{figure*}[ht]
  \begin{subfigure}[b]{0.25\linewidth}
    \includegraphics[width=0.95\linewidth,height=1.5in]{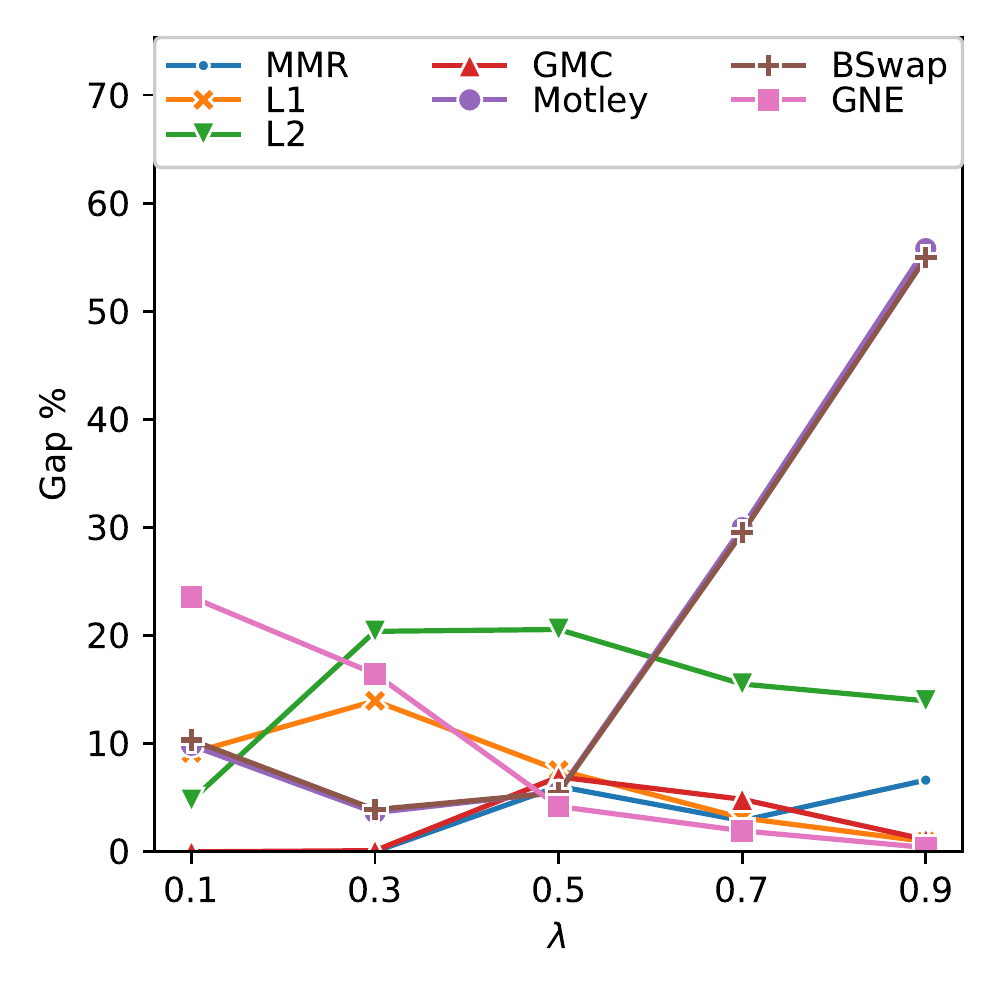}
    \caption{arbitrarily generated}
  \end{subfigure}
  \begin{subfigure}[b]{0.25\linewidth}
    \includegraphics[width=0.95\linewidth,height=1.5in]{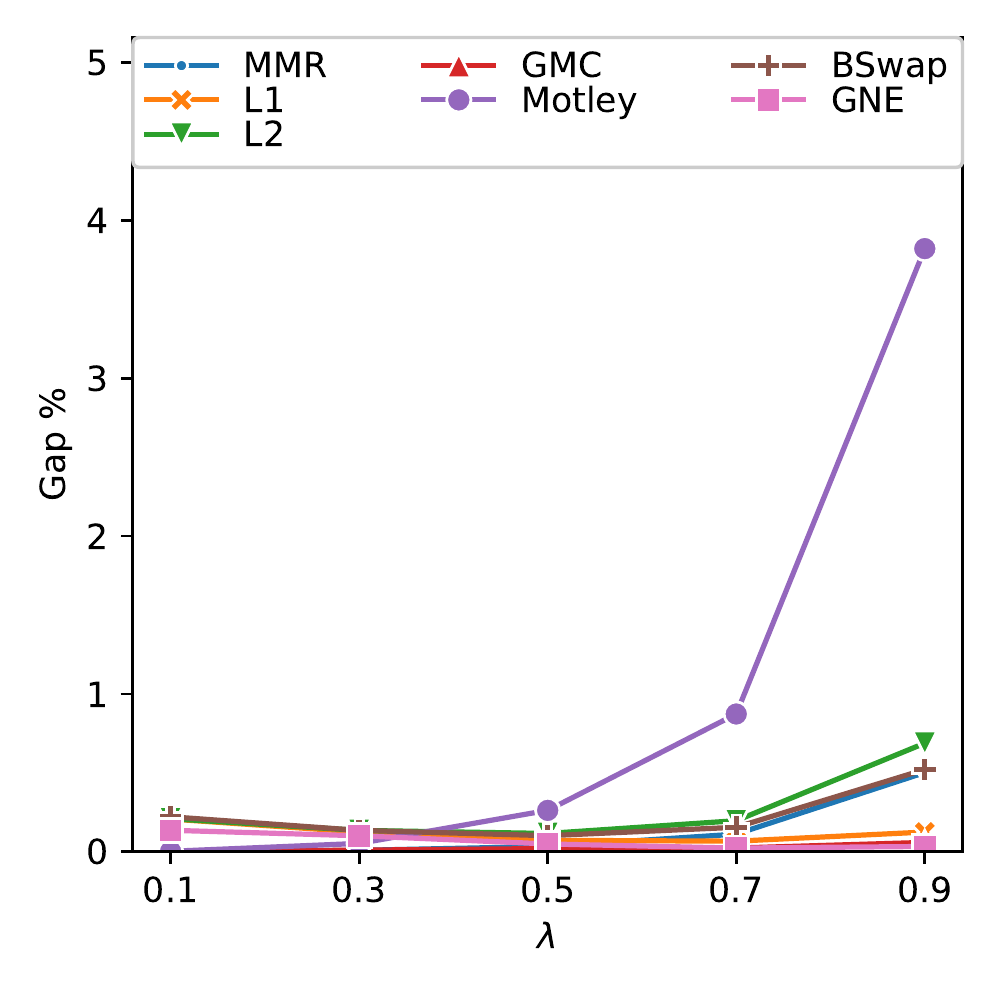}
    \caption{\emph{reuters}}
  \end{subfigure}
  \begin{subfigure}[b]{0.25\linewidth}
    \includegraphics[width=0.95\linewidth,height=1.5in]{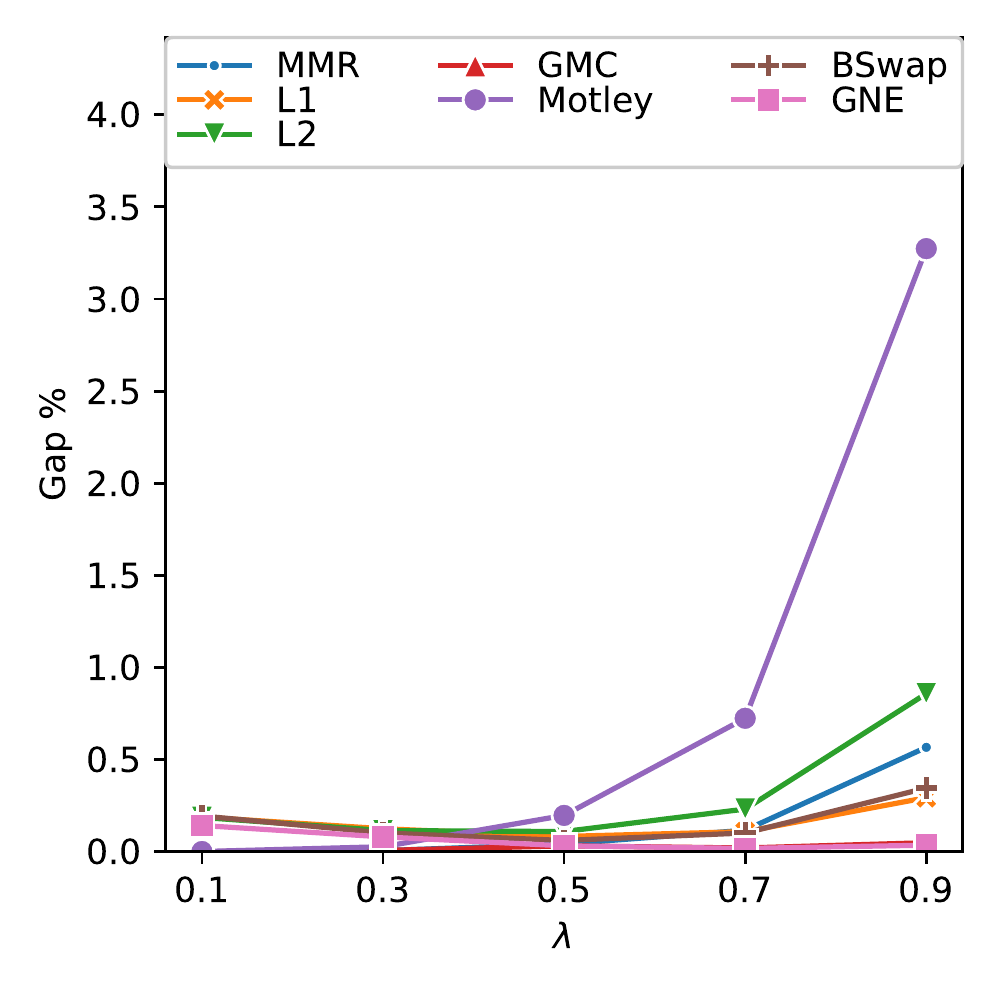}
    \caption{\emph{dblp}}
  \end{subfigure}
  \begin{subfigure}[b]{0.25\linewidth}
    \includegraphics[width=0.95\linewidth,height=1.5in]{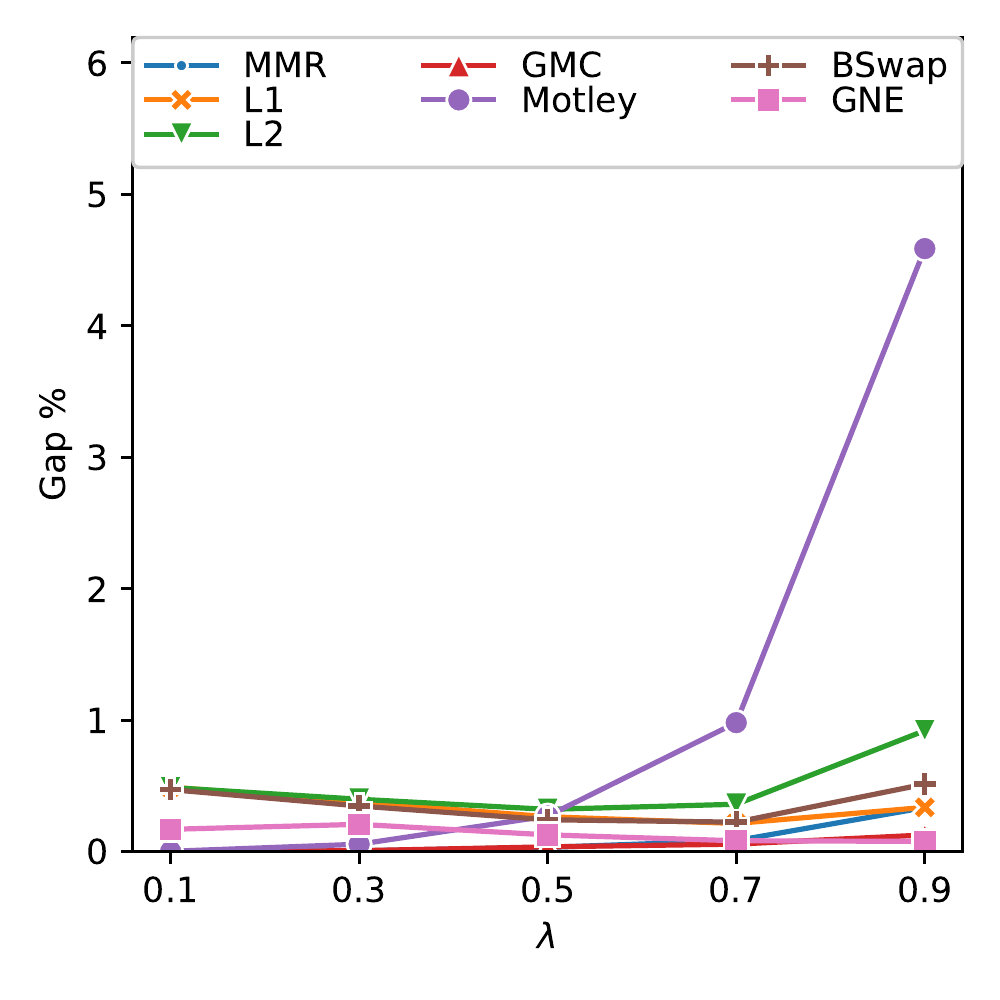}
    \caption{\emph{ads}}
  \end{subfigure}
 
  \caption{Average Gap percentage varied over values of $\lambda$}
  \label{fig:gap}
\end{figure*}
 
\begin{figure*}[ht]
  \begin{subfigure}[b]{0.25\linewidth}
    \includegraphics[width=0.95\linewidth,height=1.5in]{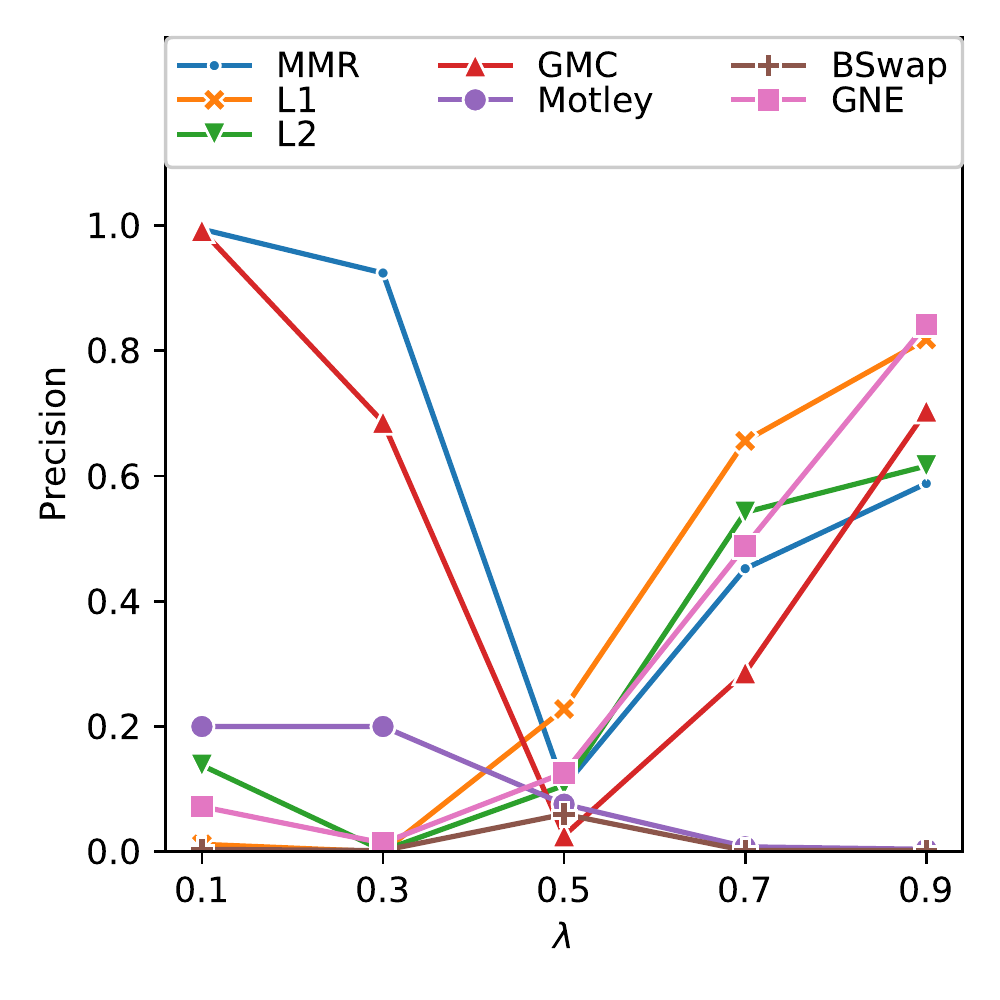}
    \caption{arbitrarily generated}
  \end{subfigure}
  \begin{subfigure}[b]{0.25\linewidth}
    \includegraphics[width=0.95\linewidth,height=1.5in]{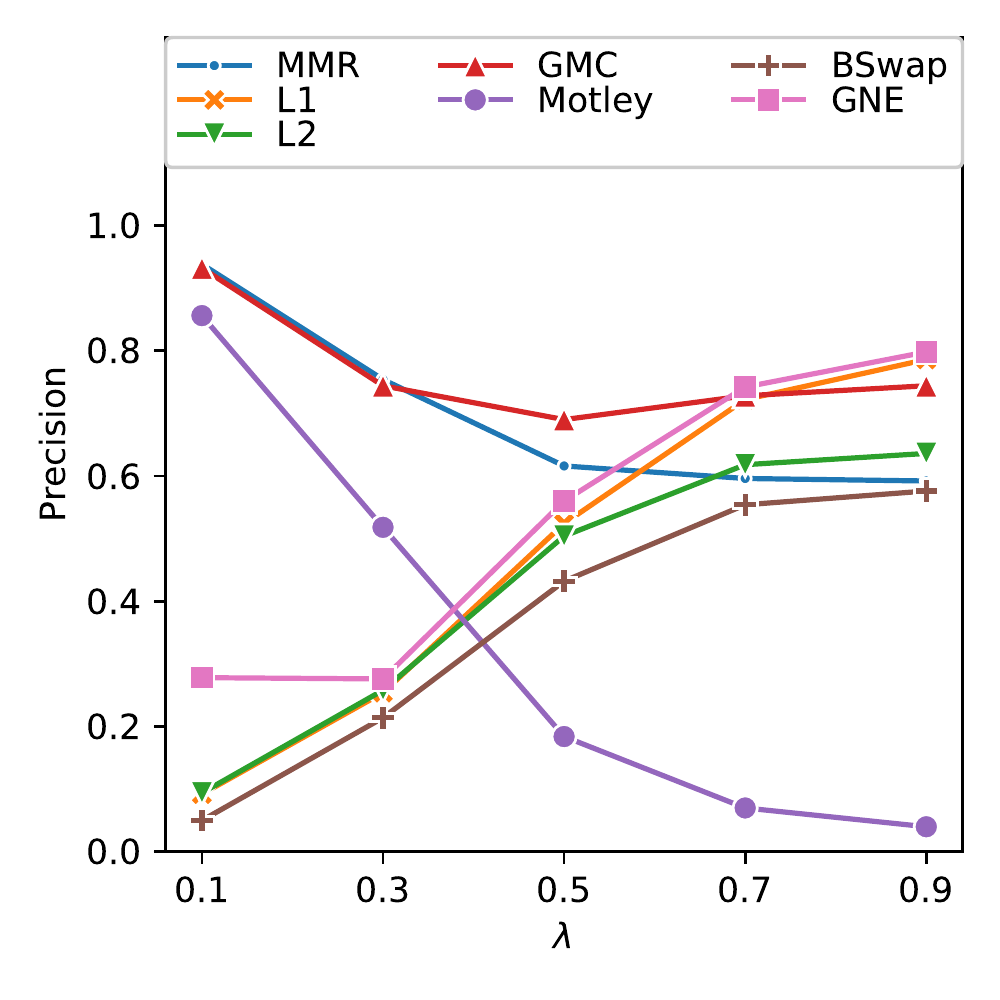}
    \caption{\emph{reuters}}
  \end{subfigure}
  \begin{subfigure}[b]{0.25\linewidth}
    \includegraphics[width=0.95\linewidth,height=1.5in]{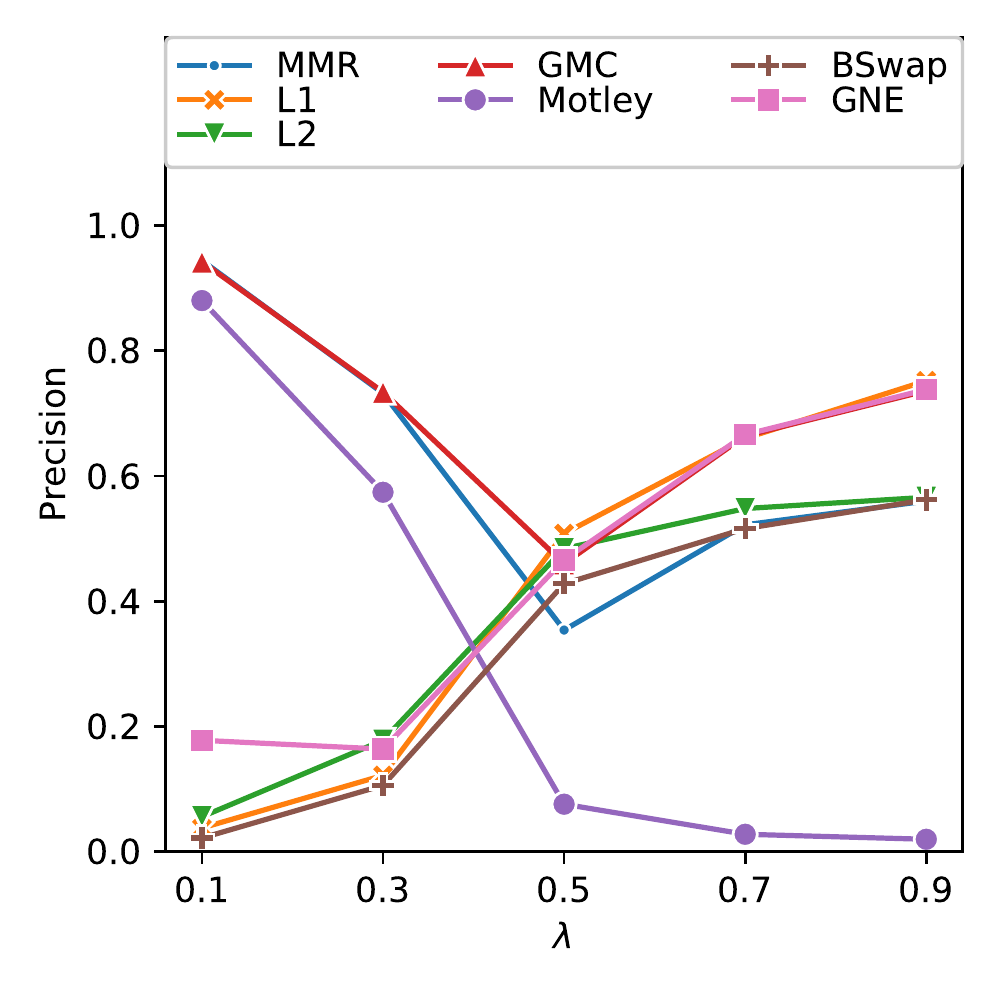}
    \caption{\emph{dblp}}
  \end{subfigure}
  \begin{subfigure}[b]{0.25\linewidth}
    \includegraphics[width=0.95\linewidth,height=1.5in]{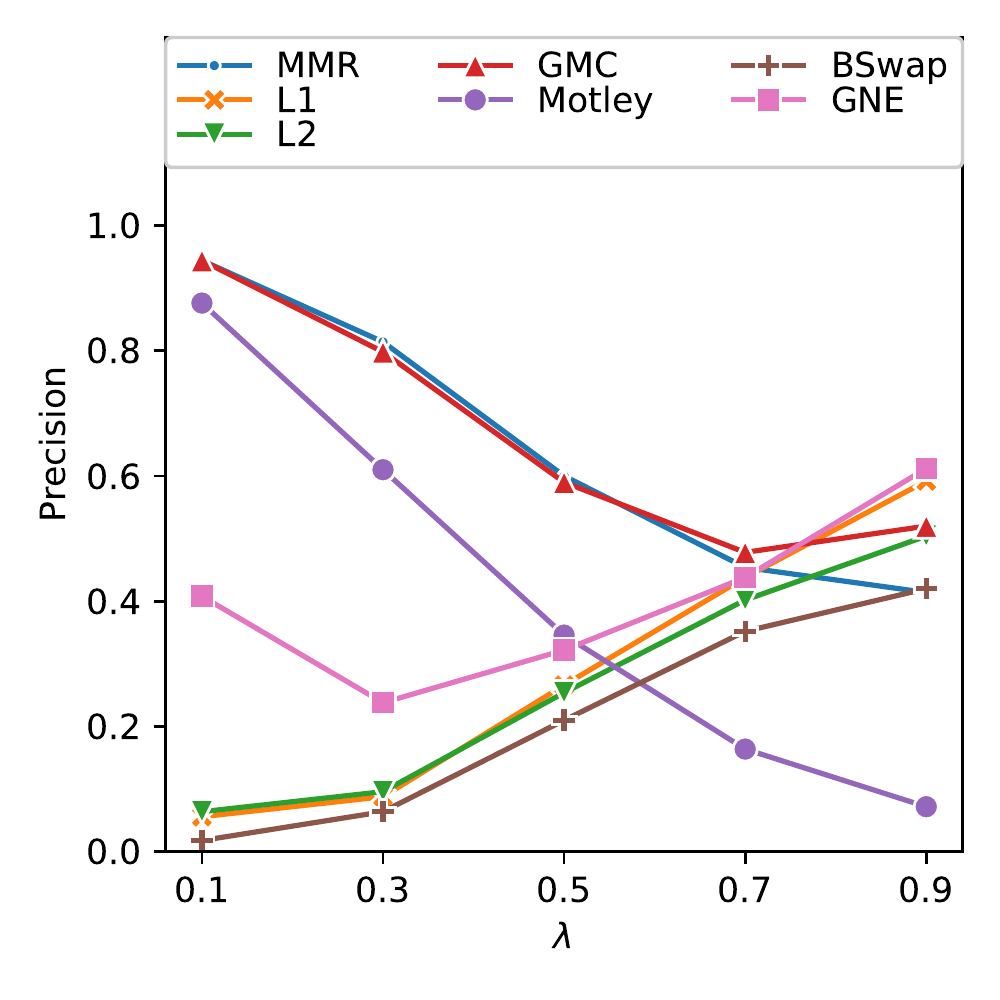}
    \caption{\emph{ads}}
  \end{subfigure}
 
  \caption{Average Precision varied over values of $\lambda$}
  \label{fig:prec}
\end{figure*}

\section{Experiments and Results}
We used the following algorithms and datasets for the purpose of comparison
\begin{itemize}
\item We use Bswap \cite{Yu09ittakes}, Motley \cite{Jain03providingdiversity}, MMR \cite{Carbonell:1998:UMD:290941.291025}, GMC and GNE \cite{DBLP:conf:icde:VieiraRBHSTT11} as baselines for both $l_2$ and $l_1$ relaxations. 
\item We evaluate all the methods on four datasets - two publicly available datasets: \href{http://kdd.ics.uci.edu/databases/reuters21578/}\emph{reuters} and \href{https://dblp.uni-trier.de/xml/}\emph{dblp}, logs from a popular search engine constituting \emph{advertising (ads)} dataset and finally on an arbitrarily generated dataset.
\end{itemize}

We randomly select 100 queries for each dataset and use a standard NGS implementation to retrieve the universal set $U$. For all datasets, we use a common feature representation of 128 dimensions based on an in-house implementation of deep learning based embedding. Note that both these steps are common for all algorithms and the proposed approach is directed at the optimization, thus the choice of the NGS implementation and the deep learning-based embedding doesn't matter. To show this, we simulate all the methods on an arbitrarily generated dataset. Wherein, we sample $n$ points randomly from a unit disc with origin as the query and supply this to every algorithm being evaluated.

To compare every algorithm, we measure the relative gap with the optimum, defined as
\begin{align}
\mbox{gap} = \frac{\lvert F(q,S^{a})-F(q,S^{*})\rvert}{F(q,S^{*})}
\end{align}
where, $S^{*}$ is the optimal set obtained by an exhaustive search and $S^{a}$ is the set obtained by the algorithm under consideration.
Similar to \cite{DBLP:conf:icde:VieiraRBHSTT11}, precision is defined as
\begin{align}
\mbox{precision} = \frac{\lvert S^{a} \cap S^{*}\rvert}{\lvert S^{*}\rvert}
\end{align}
and measures the intersection between the optimal candidate set and the algorithm's obtained set. For each query, we compute both the measures and average them across the 100 queries.

As seen in the graphs for the average gap (\ref{fig:gap}), for all datasets $l_1$ has slightly better or comparable performance as compared to all the baseline algorithms. While $l_2$ is inferior to the state-of-the-art in terms on objective but it has superior performance in terms of computational complexity. Note that $l_2$ problem is essentially an eigen value problem for a symmetric matrix and thus has efficient implementations. For $l_1$ optimization we use a standard convex optimization library and its complexity is $\mathcal{O}(n^{4.5})$ \cite{luo2010semidefinite}. The precision curves in Figure 2 for different values of $\lambda$ show that $l_1$ and GNE are comparable in higher $\lambda$ region where they outperform all the other methods. $l_2$ follows similar trend as $l_1$ and outperforms MMR in the higher $\lambda$ regions. In most of the simulations, we observed that the performance of MMR and other heuristic algorithms are strongly dependent on the distance function and the distribution of the points in the datasets. We note that the precision of $l_1$ relaxation substantially improves for higher $\lambda$ outperforming other methods whereas for the lower $\lambda$ it suffers. However, for the lower $\lambda$ the gap with optimum for $l_1$ relaxations is within 10\% for all cases.


\section{Conclusion}
In this paper, we address the problem of finding the most diverse yet relevant set of keywords/phrases for a given user query. We propose efficient and scalable convex approximations for the same. As shown in the results section the proposed framework performs fairly well on real-world datasets compared to the state-of-the-art.



\bibliographystyle{ACM-Reference-Format}
\bibliography{sample-base}

\end{document}